\documentclass[amsmath,amssymb]{revtex4}
\usepackage{graphicx}

\def \be{\begin{equation}} 
\def \ee{\end{equation}} 
\def \bea{\begin{eqnarray}} 
\def \eea{\end{eqnarray}}

\def \bfe{{\bf e}}

\def \bS{{\bf S}} 
 
\def \bL{{\bf L}} 
 
\def \cH{{\cal H }}

\def \bL{{\bf L }}

\def \cP{{\cal P}} 
\def \cO{{\cal O}} 
\def \cM{{\cal M}} 
\def\a{{\alpha}}

\def\e{{\epsilon}}

\def\W{{\Omega}} 
\def\D{{\Delta}}

\def\ket#1{{\,|\,#1\,\rangle\,}} 
\def\bra#1{{\,\langle\,#1\,|\,}} 
\def\braket#1#2{{\,\langle\,#1\,|\,#2\,\rangle\,}}

\def\nd{{^{\vphantom{\dagger}}}} 
 
\def \yBCO6{{ YBa$_2$\-Cu$_3$\-O$_{7-\delta}$ }} 
\def \yBCO6{{ YBa$_2$\-Cu$_3$\-O$_{6.6}$ }} 
\def \yBCO6x{{YBa$_2$\-Cu$_3$\-O$_{6+x}$}}

\begin{document}

\title{Computing Effective Hamiltonians of Doped and Frustrated Antiferromagnets By Contractor Renormalization}

%\classification{75.10.Jm, 75.10.Hk, 75.30.Ds}
%\keywords{High Tc Superconductivity, Renormalization, Frustration, Quantum Magnetism, Hubbard model}

\author{Assa Auerbach}
\address{Physics Department, Technion, Haifa 32000, Israel}

\begin{abstract}
 A review of the Contractor Renormalization (CORE)  method, as a systematic derivation of the low energy effective hamiltonian,
is given, with emphasis on its differences and advantages over traditional perturbative
(weak/strong links) real space RG. For the low energy physics of the 2D Hubbard model, we derive the plaquette bosons (projected SO(5))
model which connects the microscopic model to phases and phenomenology of high-T$_{\rm c}$ cuprates. For the
S = 1/2 Pyrochlore and Kagom\'e antiferromagnets, the  effective hamiltonians predict spin-disordered,
lattice symmetry breaking, ground states with a large density of low energy singlets as  
found by exact diagonalization of small clusters.
 \end{abstract}
\maketitle

%%%%%%%%%%%%%%%%%%%%%%%%%%%%%%%%%%%%%%%%%%%%
%% MAINMATTER
%%%%%%%%%%%%%%%%%%%%%%%%%%%%%%%%%%%%%%%%%%%%

\section{Relief From Strong Frustration}
\begin{figure} 
 \label{fig:frust}
\includegraphics[height=.3\textheight]{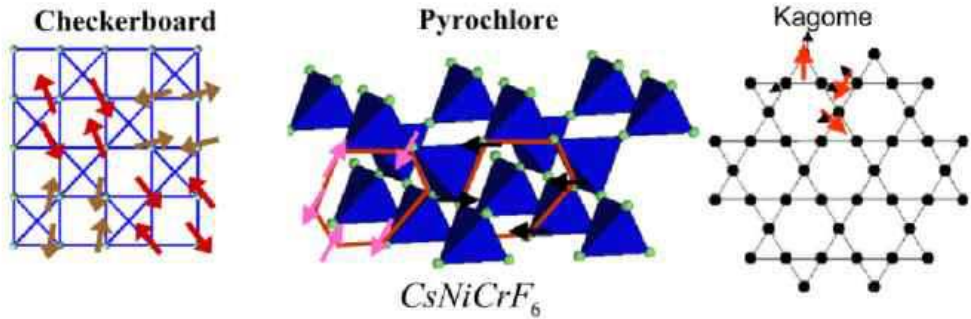}
  \caption{Strongly Frustrated Quantum Antiferromagnets treated by CORE. Red arrows denote {\em classical} spin directions.
  These can collectively rotate freely in the classical ground state manifold, rendering  a poorly controlled spin wave expansion.}
\end{figure}

Frequently, interesting models of condensed matter systems involve strong local frustration. 
For example: the Heisenberg antiferromagnet given by
\begin{equation}
H=J\sum_{\langle ij\rangle} \bS_i\cdot\bS_j\label{eq:heisenberg},
\end{equation}
where ${\langle ij\rangle} $ are nearest neighbor bonds on  lattices depicted in Fig.\ref{fig:frust}.
The classical (infinite  spin size) groundstates of the Pyrochlore and Kagom\'e lattices, are known to exhibit macroscopic (exponential in system size) 
degeneracy, which can be lifted by quantum fluctuations.

At low enough temperatures, one expects the third law of thermodynamics to 'kick in' and that quantum fluctuations will choose a particular ground state.
That ground state may, or may not, break spin rotational symmetry. 
However, sorting it out by  semiclassical expansions such as spin wave theory, is a poorly controlled endeavor.  In addition, numerical methods
generally suffer from finite size limitations, and/or minus signs in quantum Monte Carlo simulations.

Frustration causes fierce competition between nearly degenerate variational states, and equally plausible mean field theories. Hence the phase diagram of
such models are often a source of intense controversies.
We advocate that such problems  are best attacked by the 'divide and conquer' approach within a systematic real-space renormalization  scheme.
The physical analogy is the use of nucleons, and then atoms, to treat the low energy correlations of the standard model. 
It is obvious, that questions in chemistry, such as the relative stability of molecules, are better resolved using effective interactions between 
atoms rather than  by variational approximations on the high energy ('microscopic') interactions.

Returning to condensed matter physics, we have adopted the approach invented by  Morningstar  and Weinstein  called 
Contractor Renormalization (CORE)\cite{Morn96}
to treat the square lattice Hubbard model\cite{Altman02}, and several problems of frustrated quantum antiferromagnets\cite{Berg03, Budnik04}.
Other groups have also  applied CORE  to spin ladders \cite{PS97}, t-J ladders\cite{Cap02}, and frustrated antiferromagnets \cite{Cap04}.

The  essence of CORE is that the microscopic lattice Hamiltonian is   mapped onto an 
effective Hamiltonian which acts on sites of a superlattice, within a lower energy Hilbert space, as we shall review below.
%%%%%%%%%%%%%%%%%%%
%%%%%%%%%%%%%%%%%%%%%%%%%%%%%%
After an effective hamiltonian is found numerically, and represented in terms of familiar second quantized operators
(bosons, fermions, pseudospins etc.)  the remaining task is to determine its  ground state and excitations. This
could be carried out in different ways. If the effective model is still highly frustrated (as we shall find for the pyrochlore case),
the CORE method could be reiterated. If the effective model turns out to be 'simple', that is to say: apparently unfrustrated, it naturally  lends itself to  
variational approaches, and quantum Monte-Carlo methods (as was done for the Projected SO(5) theory of the square lattice Hubbard model\cite{pSO5,pSO5-num}, 
and the Quantum Clock model of the Kagom\'e \cite{Budnik04}.)

We shall see that if the effective interactions produced by CORE were calculated to all ranges, upto the size of the full lattice, 
the resulting effective hamiltonian would reproduce the {\em exact} spectrum
of the original Hamiltonian. However, this in itself does not yield saving  of numerical effort. The success of CORE relies on
a {\em rapid decay of the effective interactions with range}. This range, which is derived from the numerical convergence tests, describes the physical
{\em coherence} length of the effective degrees of freedom used to describe thee effective hamiltonian: e.g. bosonic hole pairs, for the square lattice Hubbard model,
or  pseudospins for the local singlets of the pyrochlore model.
Therefore, a proper choice of effective degrees of freedom is helpful  for rapid convergence. 
 
\section{CORE}
CORE is a {\em non-perturbative} block-spin renormalization, which uses exact diagonalizations to extract the effective
interactions.

\begin{enumerate}
\item {\it Defining the reduced Hilbert Space}. We first choose the elementary blocks which cover the lattice.
(See Fig.\ref{fig:blocks} for illustration for a square lattice).
In order to preserve as much as possible the lattice symmetries of the original model (a choice of covering always breaks
some translational symmetry), an optimal choice would be blocks which have the original  rotational symmetries: such as plaquettes in a square lattice,
triangles in the Kagom\'e and triangular lattices, and teterahedra in the pyrochlore lattice.

We diagonalize $\cH$ on a single block and truncate all states above 
a chosen cutoff energy. This leaves us with the lowest $M$ states $\{\ket{\a}\}_1^M$.
The reduced lattice Hilbert space is spanned by 
tensor products of retained block states $\ket{\a_1,\ldots,\a_N}$.
A case in point is the Hubbard model spectrum on a plaquette, which for the 
half filled case has 70 states (see Fig.\ref{fig:hub-plaq}). We truncate 66 states and keep the 
ground state and lowest triplet, i.e. $M=4$. 
Thus, the Hilbert space is considerably reduced at the first step. The retained states are in essence the {\em 'atoms'} of the new effective Hamiltonian. The next task is to find their
effective mass and interactions by calculating the intersite interactions.
  %%%%%%%%%%%%%%%%%%%
\begin{figure} 
 \label{fig:blocks}
\includegraphics[height=.3\textheight]{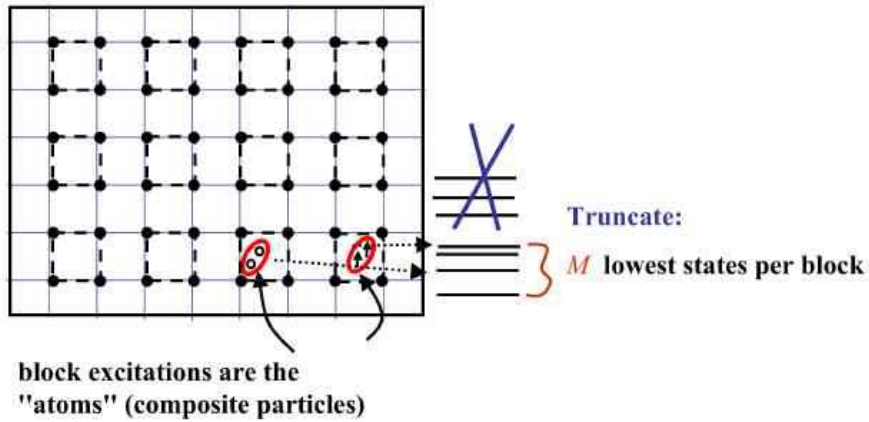}
 \caption{Covering the square lattice with plaquettes as elementary blocks. The reduced Hilbert space is defined as the
 tensor products of the lowest $M$ states in each block.}
\end{figure}
%%%%%%%%%%%%%%%%%%%%%%%%%%%%%%%%%%%%

 \item {\it The Renormalized Hamiltonian of any  cluster}. The 
reduced Hilbert space on a given connected cluster of $N$ 
blocks is of dimension $\cM=M^N$.   (See Fig. \ref{fig:hren} for an illustration for N=3).
We diagonalize $\cH$ on the cluster and obtain 
the lowest $\cM$ eigenstates and energies: $(\ket{n},\e_n)$, 
$n=1,\ldots,\cM$. The wavefunctions $\ket{n}$ are projected on the 
reduced Hilbert space and their components in the block basis 
$\ket{\a_1,\ldots,\a_N}$ are obtained. The projected states $\psi_n$ are 
then Gramm-Schmidt orthonormalized,  starting from the ground 
state upward. \be \ket{\tilde\psi_n}={1\over {Z_n}}\left( 
\ket{\psi_n}-\sum_{m<n}\ket{\tilde\psi_m} 
\braket{\tilde\psi_m}{\psi_n}\right), \label{gs} \ee where $Z_n$ is 
the normalization. The renormalized Hamiltonian is defined as 
\be 
\cH^{ren}\equiv \sum_{n}^{\cM} 
\e_n\ket{\tilde\psi_n}\bra{\tilde\psi_n},
\label{Hren}
\ee 
which ensures that 
it reproduces  the lowest $\cM$ eigenenergies exactly. 
   %%%%%%%%%%%%%%%%%%%
\begin{figure} 
 \label{fig:hren}
\includegraphics[height=.3\textheight]{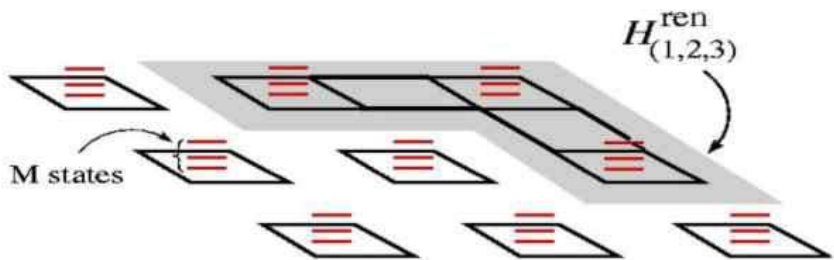}
 \caption{A Cluster of 3 blocks, defining the effective Hamiltonian $\cH^{ren}$ of range 3.}
\end{figure}
%%%%%%%%%%%%%%%%%%%%%%%%%%%%%%%%%%%%

Representing $\cH^{ren}$ in the real space block basis 
$\ket{\a_1,\ldots,\a_N}$, defines the (reducible) inter-block 
couplings and interactions. 

\item {\it Cluster expansion}. We define connected $N$ point 
interactions as: 
 \be h_{i_1,\ldots,i_N}=H^{ren}_{ 
\langle{i_1,\ldots,i_N\rangle}}-\sum_{\langle{i_1,\ldots,i_N'\rangle}} 
h_{i_1,\ldots,i_N'}, \label{hcon} \ee 
where the sum is over 
connected subclusters of $\langle i_1,\ldots, i_N\rangle$. The 
full lattice effective Hamiltonian can be expanded as the sum 
\be 
\cH_{eff}=\sum_i  h_i + \sum_{\langle ij\rangle}h_{ij} 
+\sum_{\langle ijk\rangle}h_{ijk} + ... \label{c-expansion} 
\ee 
 
$h_i$ is simply a reduced single block hamiltonian. $h_{ij}$ 
contains nearest neighbor couplings and corrections to the 
on-site terms $h_i$. $h_{ijk}$ contains three site couplings and 
so on. $h_{i_1,\ldots,i_N}$ will henceforth be called {\em range-N interaction}. 
We expect on physical grounds that for a proper choice of a 
truncated basis, range-N interactions will decay rapidly with 
$N$. This expectation  needs to be verified on a case by 
case basis. 

 \end{enumerate}

In general, there is no {\em a priori} quantitative estimation of the 
truncation error. Nevertheless, if it decays rapidly with  interaction range, 
we deduce that there is a short {\em coherence length} related to our local degrees of freedom, 
e.g. in our case the hole pair bosons and the triplets (bound states of two spinons). 

\subsection{Comparison to perturbative real-space RG}
Perturbative real-space Renormalization of Quantum Many-Body systems is carried out  in either the Lagrangian or Hamiltonian
formulation. The Lagrangian renormlization involves integrating out of the path integral high wavevector  and frequency modes $\psi_{\rm high}$. For example, in a model
with point interactions of strength $g$, the renormalization is carried out by  expanding the exponential in powers of $g$:
models
\bea
L^{\rm  pert}[\psi_{\rm low}] &=&  -\ln {1\over Z} \int {\cal D} \psi_{\rm high} \exp\left( -\psi^* L^{(2)} \psi - g|\psi|^2 \right)\nonumber\\
&\approx&   \psi^*(x,\tau)\left(  L^{(2)}+\Sigma(\tau-\tau')\right) \psi(x',\tau') + g^{\rm ren} |\psi|^2 + \cO(g^4)  
\eea

This formulation always truncates  higher order terms in $g$ (loops). It also necessarily
introduces time-retarded interactions. This procedure usually  results in a Lagrangian which  similar to the microscopic one but with renormalized coupling constants.
This allows an iterative renormalization  (the renormalization group). However time retardation, if not neglected, divorces the Lagrangian formulation from the 
operator Hamiltonian formulation.
 
The alternative is a perturbative Hamiltonian renormalization scheme. However, the traditional (non CORE) approach does not avoid the limitations of perturbation theory
and time retardation.  
First one  the Hamiltonian is separated into block terms ($H_0$) and inter-block interactions ($H'$).
\be
H=H_0+H'
\ee
The second step is to write the effective two-site hamiltonian in terms of a Brillouin-Wigner   (BW) perturbation theory in the same reduced Hilbert space 
as CORE given by by the tensor products
of the truncated  block states $|\alpha_i\rangle$.
\be
\cH^{\rm BW} =   H_0 + H'  +  
H' \sum_{n=1}^\infty \left({1-P_0\over E-H_0}H'\right)^n    
\ee
where \be
P_0 = \prod_i \sum_{\alpha_i=1}^\cM |\alpha_i\rangle\langle \alpha_i |.
\ee
The expansion for $\cH^{\rm BW}$ contains intercluster interactions of all ranges, and the sizes of the terms is controlled by $H'/\Delta$
where $\Delta$ is a typical gap energy in the spectrum of $H_0$. The appearance of $E$ inside $\cH^{\rm BW}(E)$  is
 a signature of time retarded interactions. It
means that the spectrum is not  given by the eigenvalues of $\cH^{\rm BW}(E_0)$ for any choice of $E_0$!

In summary, 
CORE  has two major advantages over traditional  perturbative  real-space
renormalization schemes:
\begin{enumerate}
\item CORE  is {\em not} an  expansion  in weak/strong bonds between block-spins.  Its convergence does not necessarily 
depend on existence of a large gap to the discarded  states of the Hilbert space.
\item CORE is based on an {\em exact} mapping from the original Hamiltonian to an effective Hamiltonian, whose truncation error can be 
estimated numerically. Non-Hamiltonian retardation effects are avoided.
\end{enumerate}

\section{Square lattice Hubbard Model}
\begin{figure} 
 \label{fig:hub-plaq}
\includegraphics[height=.5\textheight]{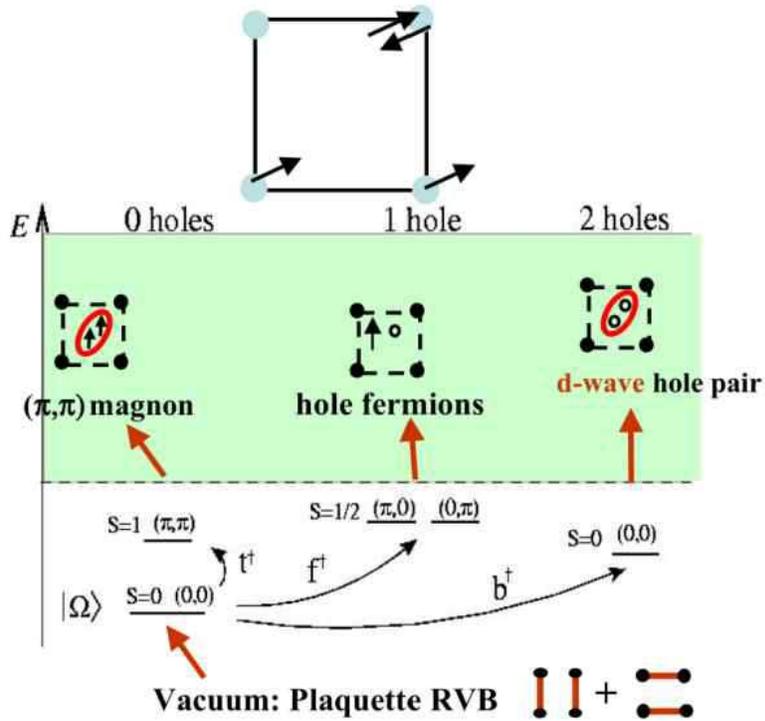}
  \caption{The low spectrum of the four site Hubbard model.  Eigenstates are labeled 
by total spin $S$ and plaquette momentum $Q_x,Q_y=0,\pi$. The undoped  ground state is the vacuum, and the excitations are
labeled by boson and fermion operators acting on that vacuum.}
\end{figure}

An important interacting many-body model, especially in the context of high temperature superconductors,
 is the square lattice Hubbard model given by
 \be 
\cH=-t\sum_{\langle i j\rangle,s}^{sl} 
\left( c^\dagger_{is} c^\nd_{js} + \mbox{H.c}  \right)   +U\sum_i 
 n_{i\uparrow} n_{i\downarrow} , \label{HM} 
 \ee  
 where $c^\dagger_{is}, n_{i s}$ are electron creation and number operators at site $i$ on the 
 square lattice. Following the CORE procedure we choose to cover the square lattice by plaquettes (as in Fig.\ref{fig:blocks}).
 The low spectrum  of 2, 3, and 4 electrons (2, 1 and no holes respectively) is depicted in Fig.\ref{fig:hub-plaq}.

%%%%%%%%%%%%%%%%%%%
%%%%%%%%%%%%%%%%%%%%%%%%%%%%%%
The ground state of the 4-site Hubbard model  at half filling ($n_e=4$)  is called $\ket\W$. 
which corresponds at large $U/t$ to the resonating valence bonds (RVB) ground state of the 
Heisenberg model plus small contributions from doubly occupied 
sites.    
The product state  $\ket\W 
=\prod_i^{plaq}\ket\W_i$, is our vacuum state for the full lattice, upon which Fock states 
can be constructed using second quantized boson and fermion 
creation operators. 

The  magnons are defined by the undoped plaquettes which are in the lowest  triplet of $S=1$ states. 

The hole pair state at ($n_e=2$)   is described by 
 \bea 
b^\dagger_\alpha \ket\W &=& 
{1\over   \sqrt{Z_b}} \cP   c^\dagger_{(0,0) \uparrow} c^\dagger_{(0,0)\downarrow} |0\rangle 
\nonumber \\ 
&=&   {1\over  \sqrt{Z'_b}} \left(\sum_{i j}  d_{ij} 
  c_{i\uparrow}  c_{j\downarrow} +\ldots \right) \ket\W, 
\label{holepairs} \eea 
 where $d_{ij}$ is +1 (-1) on vertical (horizontal) bonds, and  $\ldots$ are higher order 
 $U/t$-dependent operators. Thus, $b^\dagger$   creates a  pair 
 with internal  $d$-wave symmetry with respect to the vacuum. For the relevant range of $U/t$, the 
state normalization is $1/3<Z_b' <2/3$. 
The important energy to note is the pair binding energy defined as 
\be \D_{b} \equiv E(0) +E(2)-2E(1) 
\label{bind} \ee 
where $E(N_h)$ is the ground state of $N_h$ 
holes.    In the range $U/t \in (0,5)$, it is bounded by 
 $-0.04 t< \D_b<0$.   It has been well appreciated that the Hubbard,  t-J and even 
 CuO$_2$ models 
 have pair binding in finite  clusters starting with one plaquette.
 
 A d-wave superconducting state  can be written as the coherent state 
\be 
\Psi^{d-scF}\equiv \prod_i^{plaq} (\cos\theta +   \sin\theta e^{i\varphi}  b^\dagger_i)\ket\W    , 
\label{dSC} 
\ee 
with the superconductor order parameter 
\be 
\langle \Psi | d_{ij} c_{i\uparrow} c_{j\downarrow}   |\Psi\rangle  = 
\sqrt{Z_b'} e^{i\varphi}\sin\theta\cos\theta. 
\label{scop} 
\ee 
Both the triplets and the hole pairs are 'bosonic states', which can be represented by boson creation operators
acting on the RVB vacuum. They do not carry a negative sign under exchange. 

The single hole (3 electrons) ground states  are fermions. Since they are slightly higher  in energy 
than the hole pair states,  we truncate the spectrum below them (at our peril, of course!), for the sake of deriving
a purely bosonic effective hamiltonian, with hopefully rapidly decreasing interactions at long range.

\subsection{The Plaquette-Boson, (Projected SO(5)) Model}
We present the  CORE calculations to 
range-2 boson interactions, while projecting out the fermion states.  This required 
a  modest numerical diagonalization effort of the Hubbard model on up to 8 site clusters. 
The resulting range-2  Plaquette  Boson  (PB) model can be separated into 
bilinear  and quartic (interaction) terms: 
\be 
 \cH^{\rm pb}=\cH^b[b] +\cH^t[t]+\cH^{int}  [b,t] 
 \label{H4B} 
 \ee 
where the bosons obey local hard core constraints 
 \be 
b^\dagger_i b_i +\sum_\alpha t^\dagger_{\alpha i} t_{\alpha i} \le 1 
\ee 
The bilinear energy terms are 
 \bea 
 \cH^{\rm b} &=& (\epsilon_b - 2\mu) \sum_i b^\dagger_{i} 
b^\nd_{i} - J_b \sum_{\langle ij\rangle} \left( 
b^\dagger_{i} b^\nd_{j}+\mbox{H.c.}\right) \nonumber\\ 
\cH^{t} &=&  \epsilon_t \sum_{i\alpha  } t^\dagger_{ \alpha i} 
t^\nd_{ \alpha i} - {J_t\over 2} \sum_{\alpha \langle ij\rangle} 
(t^\dagger_{\alpha i}t^\nd_{\alpha j} + \mbox{H.c.})  \nonumber\\ 
&&~~- {J_{tt} \over 2} \sum_{\alpha \langle ij\rangle} 
(t^\dagger_{\alpha i}t^\dagger_{\alpha j} + \mbox{H.c.}). 
  \label{4BM} 
\eea 
    %%%%%%%%%%%%%%%%%%%
\begin{figure} 
 \label{fig:pars}
\includegraphics[height=.3\textheight]{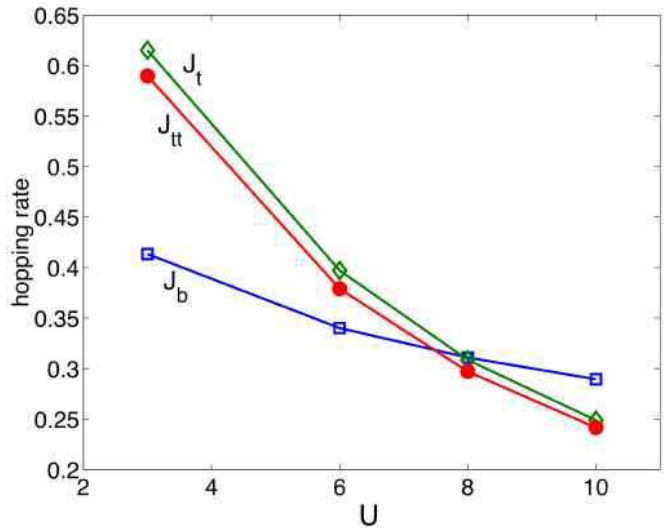}
 \caption{CORE results for the effective boson hopping rates corresponding to
 Hubbard interaction $U$  
(in units of $t$). $J_b$ is the $d$-wave hole pair hopping rate, and $J_t$ and $J_{tt}$ are magnetic energy scales defined in (\ref{4BM}).
Note near equality of hopping energies around $U/t=8$, signaling the projected SO(5) point.}
\end{figure}
%%%%%%%%%%%%%%%%%%%%%%%%%%%%%%%%%%%%

In Fig.~\ref{fig:pars} we compare the magnitudes of the magnon hoppings $J_t,J_{tt}$ and the hole pair hopping $J_b$ 
for a range of $U/t$.  The region of intersection  near $U/t=8$, 
is close to the {\em projected SO(5) symmetry 
point}. We emphasize that 
although there is {\em no quantum SO(5) symmetry} in  $H^{\rm pb}$, there is 
an approximate equality of the bosons hopping energy scales. 
This equality which was  assumed in the 
pSO(5) theory\cite{pSO5}, previously appealed 
to phenomenological considerations. Here, the equality emerges 
in a physically interesting regime of the Hubbard model and has important consequences on 
the phase diagram as was   shown in   Ref.\cite{pSO5-num}.

\subsection{CORE convergence and Coherence Length}
By diagonalizing the 12 site clusters, we have found that range 3 interactions are indeed between 1-10\%
of the range 2 interactions.
Computing range 3 interactions  $h_{123}$ and finding out whether they are significantly smaller than range 2 terms
is   important   for two main reasons: (i) This is the only way one could validate a truncation of the cluster expansion
to range 2 for further investigations of the low energy properties of the model, and (ii) a rapid decrease in effective interactions signals
a short 'coherence lengthscale' which describes the size of the effective degrees of freedom. For cuprates,
the effective size of the hole pair is of experimental importance, 
since it is  bounded by the superconducting coherence length as observed by the vortex core
size, and the short superconducting healing length near grain boundaries and defects. 
Both have been observed to be not much larger than a few lattice constants.

\section{Pyrochlores}
\begin{figure} 
 \label{fig:pyro-spect}
\includegraphics[height=.5\textheight]{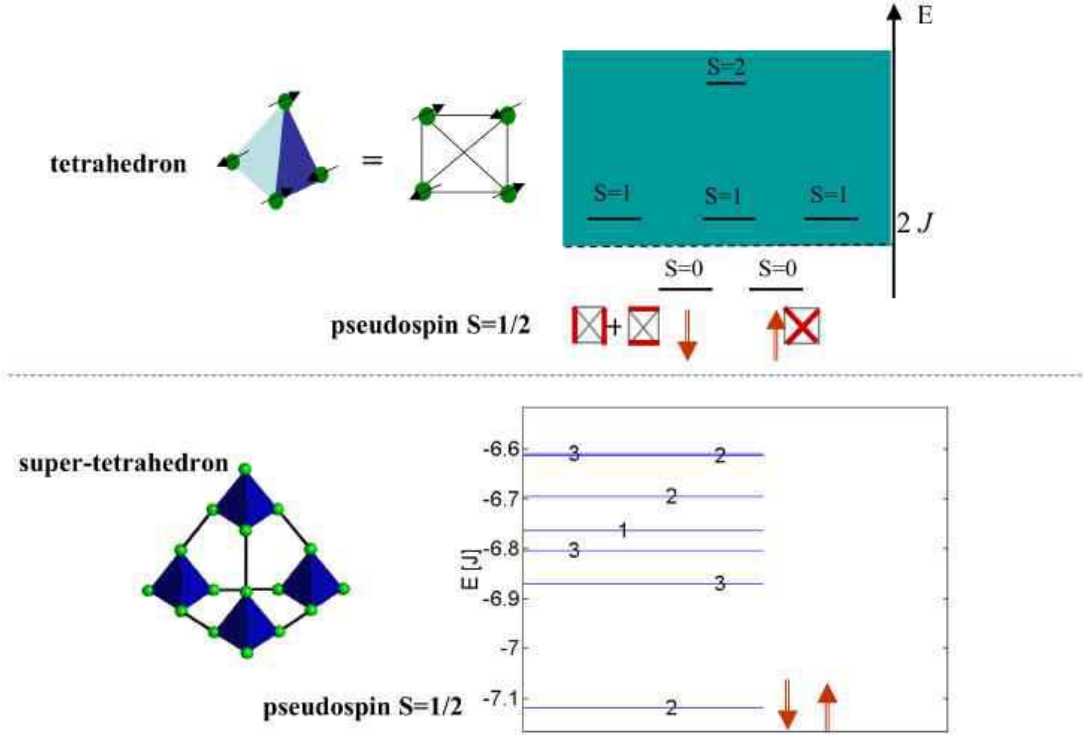}
  \caption{The low spectrum of the Heisenberg model on a tetrahedron, and a supertetrahedron.
 In both clusters the ground states are degenerate singlets which can be represented by spin half eigenstates.}
\end{figure}

The pyrochlore lattice is depicted in Fig.\ref{fig:frust}.
Depicted in Fig. \ref{fig:pyro-spect} is the spectrum of the Quantum Heisenberg model (\ref{eq:heisenberg}) on  a 4-site tetrahedron and a 16 site supertetrahedron.
Both blocks can be used to cover  the Pyrochlore lattice. Their block ground states are doubly degenerate singlets, and thus
the effective hamiltonians are readily described
by pseudospin-half operators. 
%%%%%%%%%%%%%%%%%%%

Using CORE for the tetrahedra covering  upto range 3,  we have found
\begin{eqnarray}
\label{eq:pyro1_Heff3} H^{\rm FCC} &&=  \sum_{\langle ijk
\rangle}\Bigg((J_2 ({\bf S}_i \cdot {\bf e}^{(i)}_{ijk}) ({\bf
S}_j \cdot {\bf e}^{(j)}_{ijk})-
\\
\nonumber &&J_3 (\frac{1}{2}-{\bf S}_i \cdot {\bf e}^{(i)}_{ijk})
(\frac{1}{2}-{\bf S}_j \cdot {\bf e}^{(j)}_{ijk})
(\frac{1}{2}-{\bf S}_k \cdot {\bf e}^{(k)}_{ijk})\Bigg).
\end{eqnarray}

The coupling parameters (in units of $J$) are:  $ J_2=0.1049  $,  $ J_3=0.4215
$,
  and $ {\bf e}^{(i)}_{123}, i=1,2,3$ are three unit vectors  in the x-y
plane
 whose angles $\a^{(i)}_{123}$ depend on the particular plane
defined by the triangle of tetrahedral units  $123$  as given in
table I of \cite{ts02}. The effective hamiltonian
(\ref{eq:pyro1_Heff3}) resembles the terms obtained by Tsunetsugu by second
order perturbation theory (in inter-tetrahedra couplings) \cite{ts02}. The
 classical mean field ground state is three of the four FCC
sublattices are ordered in the directions ${\bf e}(0),{\bf
e}(2\pi/3),{\bf e}(-2\pi/3)$, while the direction of the fourth is
completely degenerate.  Therefore,  classical mean field approximation for
(\ref{eq:pyro1_Heff3}) is insufficient to remove the
ground state degeneracy.  Tsunetsugu\cite{ts02} was able to
lift the degeneracy by including  spinwave fluctuations effects  which
produce  ordering at a new low energy scale.

Here we  avoid the {\em a-priori}   symmetry breaking needed for
semiclassical spinwave theory, by treating  (\ref{eq:pyro1_Heff3})   fully
quantum mechanically.  This entails a second CORE  transformation which
involves choosing  the  {\em
``supertetrahedron''}, as a basic cluster of four  tetrahedra.

 Our new
 pseudospins ${\bf \tau}_i$
are defined by the two degenerate singlet ground states of the
supertetrahedron. (This degeneracy is found for  the
  Heisenberg
model on the original lattice as well as for  the effective
 model
(\ref{eq:pyro1_Heff3})).
  These states transform as the E irreducible
representation of the tetrahedron ($T_d$)
  symmetry group, similarly to
the singlet ground states of a
 single tetrahedron.

The supertetrahedra form a cubic lattice. The effective hamiltonian (\ref{eq:pyro1_Heff3})
and the lattice geometry imply that  non-trivial effective
interactions appear only at the range of three supertetrahedra and higher.
Range three effective interactions include  two and three
pseudospin interactions, which are dominated by

\begin{eqnarray}
\label{eq:pyro2_Heff2} \mathcal H^{\rm Cubic}&=&J_1 \sum_{\langle ij
\rangle}{({\bf \tau}_i \cdot {\bf f}_{ij}) ({\bf \tau}_j \cdot
{\bf f}_{ij})}+
\\
\nonumber && J^{(a)}_2 \sum_{\langle \langle ij \rangle
\rangle}{({\bf \tau}_i \cdot {\bf f}_{ij}) ({\bf \tau}_j \cdot
{\bf f}_{ij})}+
\\
\nonumber && J^{(b)}_2 \sum_{\langle  \langle ij \rangle
\rangle}{({\bf \tau}_i \cdot ({\bf f}_{ij} \times \hat {\bf z}))
({\bf \tau}_j \cdot ({\bf f}_{ij}\times \hat {\bf z})}).
\end{eqnarray}
Here,  $\langle ~\rangle$ and $\langle \langle~ \rangle \rangle$
indicate summation over nearest- and next
nearest-neighbors, respectively. The coupling constants  are found to be
relatively small:  $J_1=0.048J$,  $J^{(a)}_2=-0.006 J$  and
$J^{(b)}_2=0.018J$ . The vectors ${\bf f}_{ij}$
depend on the vector ${\bf r}_{ij}$ connecting the two sites, and
their values are presented in Ref.\cite{Berg03}.

  %%%%%%%%%%%%%%%%%%%
\begin{figure} 
 \label{fig:pyro}
\includegraphics[height=.4\textheight]{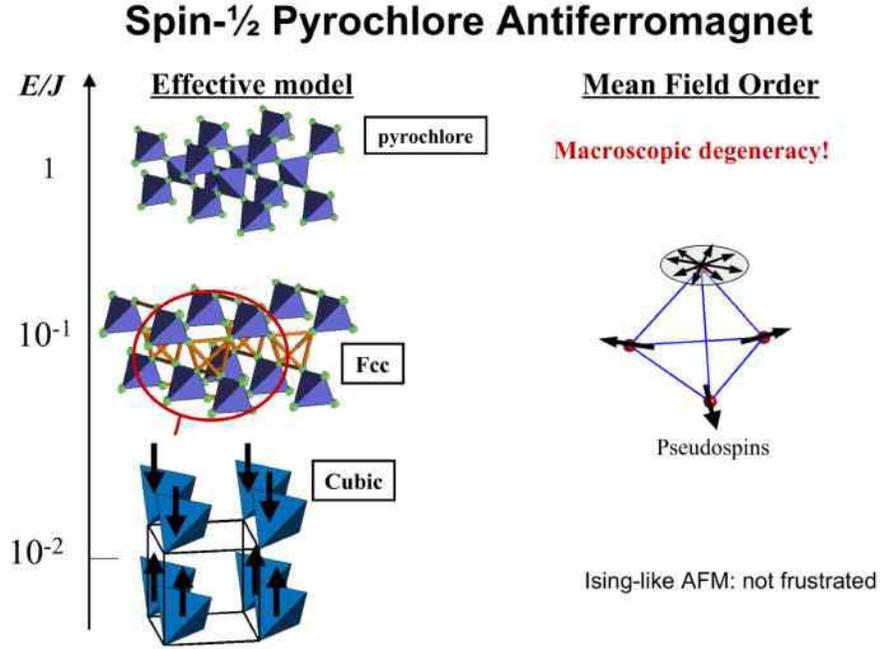}
 \caption{Two CORE steps to relieve frustration in the Pyrochlore model. The first step results in an FCC pseudospin model whose mean field solution has one  sublattice
 of completely free spins. The second CORE step results in a simple cubic model, with renormalized coupling constant of  $\sim 0.01J$, which has no residual macroscopic degeneracy.
 Its pseudospins, (which are represented by singlets on 16-site  super-tetrahedra), are antiferromagnetically correlated between neighboring planes.}
\end{figure}
%%%%%%%%%%%%%%%%%%%%%%%%%%%%%%%%%%%%

We performed classical Monte Carlo simulations using the classical (large
spin) approximation to    (\ref{eq:pyro2_Heff2}). The
ground state
 was found to choose an antiferromagnetic axis, and to be
ferromagnetic in the planes  as depicted in Fig. \ref{fig:pyro}.
It differs from  the
semiclassical ground state\cite{ts02}. The latter involves
condensation   of  high energy  states of the supertetrahedron in the
thermodynamic ground state.  Since on a  supertetrahedra
we find  a much  larger gap to these states than inter-site
coupling,  we believe they cannot condense to yield the
semiclassical ground state symmetry breaking.

To estimate the truncation error we calculated the contribution of range-4 interactions
in both stages of CORE leading to (\ref{eq:pyro1_Heff3})
and (\ref{eq:pyro2_Heff2}). Evidently, these terms are small ($<30\%$),
and most importantly,
including them does not alter the mean field solution.
\section{The Kagom\'e}
\begin{figure} 
 \label{fig:uptri}
\includegraphics[height=.3\textheight]{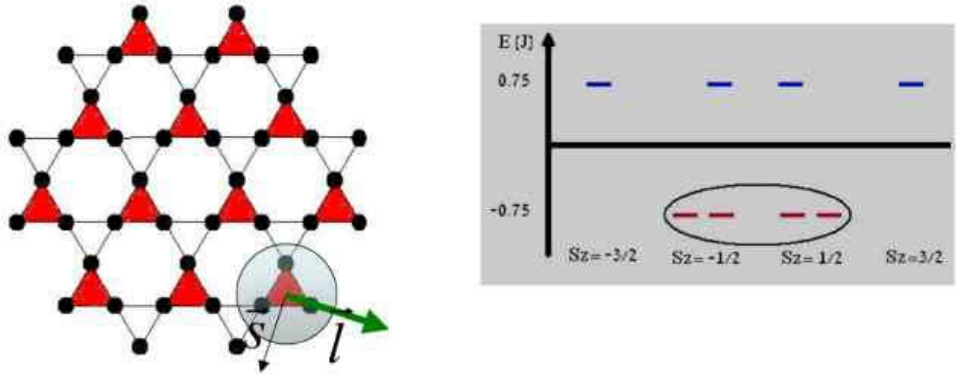}
 \caption{CORE on the Kagom\'e. Up-triangles provide four degenerate groundstates, which can represent a spin and a pseudospin of sizes half on each block.}
\end{figure}

 The Heisenberg model (\ref{eq:heisenberg}) on the Kagom\'e lattice (depicted in Fig.\ref{fig:frust})  has macroscopic degeneracy in its classical
ground state.
For the initial stage of CORE, we choose
the upward triangles covering, and a truncated basis of the four  degenerate
spin half ground states, discarding the higher $S=3/2$ states, see Fig.\ref{fig:uptri}.,
%%%%%%%%%%%%%%%%%%%
%%%%%%%%%%%%%%%%%%%%%%%%%%%%%%%%%%%%

The S-L representation of the four ground states are labeled by
$|s,l\rangle$, where $s=\uparrow,\downarrow$ is the magnetization
and $l=\Uparrow,\Downarrow$ is the pseudospin in the $z$
direction. Explicitly, in the Ising basis $|s_1 s_2 s_3\rangle $,
 \bea
|s, \Uparrow \rangle &=&    {(|s\uparrow \downarrow \rangle
 -|s\downarrow \uparrow \rangle ) \over \sqrt{2}}\nonumber\\
|s, \Downarrow \rangle  &=& {|s \uparrow \downarrow \rangle
+|s\downarrow \uparrow \rangle ) \over \sqrt{6}} - \sqrt{2\over 3}
| (-s)  s  s \rangle  \label{eq:states} \eea

The pseudospin direction in the $xz$ plane correlates with the
direction of the singlet bond, e.g. $\Uparrow$ describes a singlet
dimer on the bottom ($-\hat{z}$) edge. Thus, the $L^y$ eigenstates have
definite chiralities.

\subsection{The  SL Hamiltonian}. The effective interactions between
triangles is calculated by CORE. We note that this
approach is feasible  when two conditions are met: (i) Interaction
matrix elements  fall off rapidly with range such that the
truncation error at finite ranges is small, and (ii) the norms of
the projected eigenstates are sufficiently large for numerical
accuracy.  We have computed all range 2 and range 3 interactions,
and neglected range 4 corrections, whose expectation values were
found to be an order of magnitude smaller. At range 3, norms of
projected eigenstates were greater than 0.75, with most states
above 0.9.

The effective Hamiltonian is a Spin-Pseudospin (SL) Model on the
triangular lattice:
\begin{eqnarray}
\cH_{SL}&=&\cH_{ss}+\cH_{ll}\nonumber\\
\cH_{ss}&=&\sum_{\langle ij\rangle} \bS_i \cdot \bS_j~
\left[J_{ss}+J_{sslele}( \bL_i\cdot {\bf e}_{ij} ) \cdot (
\bL_j\cdot{\bf e}_{ji} )
\right.   \nonumber\\
& & +J_{ssll}(\bL^\perp_i\cdot \bL^\perp_j)+ J_{ssle1}(\bL_i\cdot{\bf e}_{ij} ) \nonumber\\
&& \left.+J_{ssle2}(\bL_j\cdot{\bf e}_{ji} )+J_{sslyly}\bL_i^y \bL_j^y    \right]  \nonumber\\
\cH_{ll}&=&J_{lele}( \bL_i\cdot {\bf \tilde{e}}^{ij} ) \cdot (
\bL_j\cdot{\bf \tilde{e}}_{ji} )+
J_{ll}(\bL^\perp_i\cdot \bL^\perp_j)  \nonumber\\
&&+ J_{lyly} \bL_i^y \bL_j^y   \label{Heff}
\end{eqnarray}
Here $\bL^\perp=(\bL^x,\bL^z)$, and ${\bf e}_{ij},\bfe_{ij}^l$ are
unit vectors in the $xz$ plane. $\cH_{ss}$ describes interactions
of the Kugel-Khomskii type, where the pseudospin
exchange anisotropy depends on the sites and bond directions.  For any other other bond $\langle
ij'\rangle$,  $\bfe_{ij'}$ is simply found by rotating ${\bf e}_{ij}$
by $0,\pm 120^{\circ}$ according to the O(2) rotation of $ \langle
ij \rangle \to \langle ij'\rangle$.

\subsection{Ground state and Low Excitations}
%%%%%%%%%%%%%%%%%%%
\begin{figure} 
 \label{fig:dimers}
\includegraphics[height=.3\textheight]{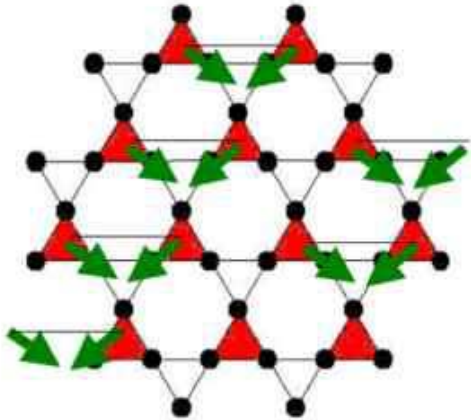}
 \caption{Variational minimum of the SL Hamiltonian. Arrows denote pseudospin directions, and solid lines denote singlet spin correlations between neighboring
 blocks. The ground state is a columnar dimer state of neighboring up-triangle singlets.}
\end{figure}
%%%%%%%%%%%%%%%%%%%%%%%%%%%%%%%%%%%%

The best variational candidate for the SL model (\ref{Heff})
are the dimer coverings of two triangle singlets, whose
correlations are defined by Fig.~\ref{fig:dimers}. The dimer
singlet states have been shown by Mila and Mambrini\cite{MM} to
span much of the low singlet spectrum in finite cluster
calculations. The variational analysis highlights the special role
of the {\em ``Dimerization Fields''}, $J_{ssle_1},J_{ssle_1}$ in
(\ref{Heff}), for the formation of local singlets. These terms
cancel under summation in all uniform states defined by $\langle
\bS_i \bS_j\rangle=\mbox{const}$. Their significant magnitude   helps to lower the energy considerably by
aligning $\bL_i$ with the anisotropy vectors $\bfe_{ij}$ to form
singlets on certain bonds and not on others $\langle \bS_i \cdot
\bS_j \rangle=-{3\over 4}\delta_{\langle ij\rangle_{dim} }$. {\em
This is a strong argument in favor of a paramagnetic ground
state.} Consequently, $\cH_{ll}$ is crucial in selecting the true
ground state among the multitude of dimer singlet coverings. We
have found that the perfectly ordered {\em columnar dimer} (CD)
state minimizes $\cH_{ll}$.  A local ``defect'' of a rotated dimer
in the CD background costs a ``twist'' energy of $+0.01$ per site. In (\cite{Budnik04}) a theory of long wavelength fluctuations about the CD state
has been investigated. The theory has a 6-fold 'clock mass term' $u_6 \cos(6\phi)$, which yields a finite  gap. 
However, quantum fluctuations renormalize down the magnitude of the clock mass to exponentially low values due to the 6 fold symmetry. This may explain the large density of 
low energy singlet excitations observed numerically\cite{singlets}, and predicts a $T^2$ temperature dependence of the specific heat due to singlet excitations.

\section{Summary}
In summary I have reviewed our group's recent applications of CORE to currently interesting problems of strong quantum frustration,
e.g.:  the 2D Hubbard model for cuprates, and the geometrically frustrated Heisenberg model. We find that the effective Hamiltonians are not necessarily simpler in form, but
in a many cases they reveal the low energy degrees of freedom as the eigenstates of small clusters. The effective interactions, if they decay rapidly in space, 
may yield less competition and frustration than in the microscopic Hamiltonian, and thus be better amenable to variational solutions.

In the computational sense, one could view CORE as an efficient algorithm to obtain the low energy physics of a large many body system, which maximally extracts its information from
exact diagonalizations of small clusters. In combination with other approaches, this provides a promising direction to disentangle the low energy physics
of strongly correlated many body problems. The method allows self-estimation of convergence, by sampling interactions of higher ranges than are retained.

\subsection{Acknowledgments}
I thank  my collaborators E. Altman, E. Berg and R. Budnik, and acknowledge fruitful discussions with S. Capponi, D. Poilblanc and R. Moessner.
I acknowledge grants from US-Israel Binational Science Foundation and the Israel Science Foundation.

%% aipguide.eps for further information.


\bibliographystyle{aipproc}   % if natbib is available

\bibliography{sample}

\IfFileExists{\jobname.bbl}{}
 {\typeout{}
  \typeout{******************************************}
  \typeout{** Please run "bibtex \jobname" to optain}
  \typeout{** the bibliography and then re-run LaTeX}
  \typeout{** twice to fix the references!}
  \typeout{******************************************}
  \typeout{}
 }

%%%%%%%%%%%%%%%%%%%%%%%%%%%%%%%%%%%%%%%%%%%
%% The following lines show an example how to produce a bibliography
%% without the help of the BibTeX program. This could be used instead
%% of the above.
%%%%%%%%%%%%%%%%%%%%%%%%%%%%%%%%%%%%%%%%%%%

\end{document}